\documentclass[sigconf]{acmart}
\AtBeginDocument{%
  }

\copyrightyear{2026}
\acmYear{2026}
\setcopyright{cc}
\setcctype{by}
\acmConference[ICMI Companion '26]{Companion of the INTERNATIONAL CONFERENCE ON MULTIMODAL INTERACTION}{October 05--09, 2026}{Napoli, Italy}
\acmBooktitle{Companion of the INTERNATIONAL CONFERENCE ON MULTIMODAL INTERACTION (ICMI Companion '26), October 05--09, 2026, Napoli, Italy}
\acmDOI{10.1145/3776591.3837060}
\acmISBN{979-8-4007-2319-3/2026/10}

\begin{document}

\title{How AI Experiences Art: Emergent Aesthetic Structure in a Self-Supervised Multimodal Embedding Space}

\author{Corey D.C. Heath}
\email{research@coreydheath.me}
\orcid{0000-0001-8902-4241}
\affiliation{%
  \institution{Independent Researcher}
  \city{Phoenix}
  \state{AZ}
  \country{USA}
}

\renewcommand{\shortauthors}{Heath}

\begin{abstract}
Aesthetics are an important part of the symbolism of artistic works. Although subjective, humans categorize art based on the emotion evoked regardless of modality. What remains underexplored is how AI models form their own aesthetic categorization of human-produced media without explicit labels or cross-modal supervision. We present a self-supervised framework that projects four modalities (text, audio, image and video) into a shared 256-dimensional embedding space and applies iterative clustering to discover aesthetic structure. We discuss the divergence between AI-generated cluster assignments and human affective register labels on a weakly supervised multimodal dataset. This work has applications in understanding how AI structures cross-modal similarity, organizing heterogeneous media collections for Retrieval-Augmented Generation (RAG), and automated data labeling.
\end{abstract}



\maketitle

\section{Introduction}
Artistic works have the ability to elicit emotion and affect from the humans who consume them. People’s tastes in art are highly personalized; however, there is general agreement on how artistic works can be categorized based on their aesthetic. Foundation AI models have been trained using human content. Our research seeks to determine how well AI understanding extends into aesthetics across different modalities without explicit human labels, and if these AI categorizations diverge from human categorization in a meaningful way.

In order to explore AI aesthetic categorization, an architectural framework was built utilizing AI models to embed multimodal content featuring text, audio, images and video. These modalities are projected into a shared space and a clustering model is used to categorize the samples. The architectural models are trained and evaluated on a weakly supervised dataset of publicly available media content. The primary contributions of this work are:
\begin{enumerate}
    \item A self-supervised framework for utilizing latent space representations for AI aesthetic self-organization of multimodal data.
    \item A divergence analysis between AI and human aesthetic topology.
\end{enumerate}

\section{Related Works}
Self-supervised learning aesthetics has been explored in previous works \cite{sheng2020revisiting, pfister2021self, yang2024self, jia2025self} focusing on photographic quality; however, in this work we are exploring affective-experiential representations in media. 

The aesthetic classification used in this project is based on the valence-arousal scale \cite{russell1980circumplex}. Picard \cite{picard1997w} established a relevant foundation for machine understanding of human emotional states. Research on emotion and affective response from images \cite{mikels2005emotional} and audio  \cite{zentner2008emotions} provide a theoretical backing of media classification based on aesthetics. Image classification based on the emotional state was implemented by Machajdik and Hanbury \cite{machajdik2010affective} using a Naive Bayes model. 

Unlike ImageBind \cite{girdhar2023imagebind}, which aligns modalities through explicitly paired training samples using images as a binding anchor, our architecture requires no-cross modal pairs. Alignment in shared space in our architecture emerges from the self-supervised clustering signal alone. 

Deep clustering methods such as DeepCluster \cite{caron2018deep} and SCAN \cite{van2020scan} extend self-supervised representation learning to cluster discovery within a single visual modality. Our framework generalizes this paradigm to four modalities simultaneously without requiring cross-modal paired sampling.

\begin{figure*}[t]
  \centering
 \includegraphics[width=\linewidth]{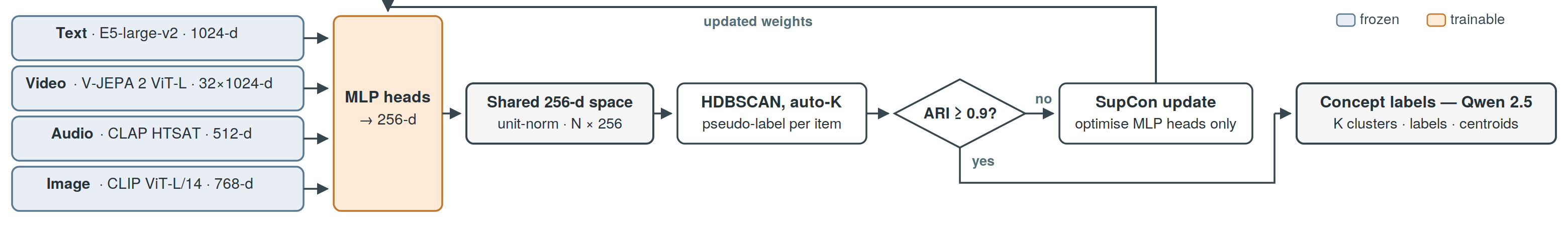}
  \caption{Architectural flow diagram depicting training process.}
  \label{fig:arch_flow}
  \Description{Block diagram showing the flow from the original sample through the backbone encoder model into the modality MLP and through the clustering process. The flow is iterative, terminating when cluster assignments are stable.}
\end{figure*}

\section{Methodology}

A weakly-supervised dataset was created for this project based on publicly available media. Data procurement was based on six categories inspired by the valence-arousal scale. The categories are elegiac, sublime, grotesque, uncanny, idyllic and pastoral. This particular phrasing was chosen for its alignment with literary and critical theory tradition. Four types of media were collected: text (consisting of poems and prose), audio, images and video. For large audio and video files, only the first 120 seconds are sampled. The data split is available in Table~\ref{tab:freq_transposed}.

\begin{table}[htbp]
  \caption{The number of samples in the weakly supervised dataset by affective register (rows) and modality (columns).}
  \label{tab:freq_transposed}
  \centering
  \begin{tabular}{l cccc}
    \toprule
    \textbf{Affective Register} & \textbf{Text} & \textbf{Audio} & \textbf{Image} & \textbf{Video}\\
    \midrule
    \textbf{Elegiac}   & 840 & 389 & 523 & 380\\
    \textbf{Sublime}   & 789 & 553 & 469 & 521\\
    \textbf{Grotesque} & 967 & 421 & 378 & 384\\
    \textbf{Uncanny}   & 706 & 333 & 408 & 380\\
    \textbf{Idyllic}   & 791 & 342 & 553 & 381\\
    \textbf{Pastoral}  & 725 & 268 & 345 & 164\\
    \bottomrule
  \end{tabular}
\end{table}

Figure~\ref{fig:arch_flow} shows a diagram of the framework architecture. The first step was to extract embedding representations of each of the samples using a specific model for each modality. The backbone models for extracting embeddings from each modality are presented in Table~\ref{tab:models}. Each model was run locally on a system using an RTX 3090 GPU.

\begin{table}[htbp]
  \caption{Models utilized to extract embeddings from each modality.}
  \label{tab:models}
  \centering
  \begin{tabular}{l l l}
    \toprule
    \textbf{Modality} & \textbf{Model} & \textbf{Citation} \\
    \midrule
    \textbf{Text}  & E5-large-v2     & Wang et al. \cite{wang2022text}\\
    \textbf{Audio} & CLAP HTSAT      & Wu et al. \cite{wu2023large} \\
    \textbf{Image} & CLIP ViT-L/14   & Radford et al. \cite{radford2021learning} \\
    \textbf{Video} & V-JEPA 2 ViT-L  & Assran et al. \cite{assran2025v} \\
    \bottomrule
  \end{tabular}
\end{table}

After extraction, each modality embedding has its own dimensional space representation. Multi-layer perceptrons (MLP) are used to project each modality into a 256-d unit-norm representation. Modality alignment happens through SupConLoss \cite{khosla2020supervised}. The training loop has two phases. In Phase 1, The 256-d output vectors for each sample are given a pseudo-label created by HDBSCAN \cite{campello2013density}. In Phase 2, the pseudo-labels are held fixed while SupConLoss trains the MLPs for 50 epochs, pulling same-cluster samples together and pushing different-cluster samples apart. Gradients flow into the MLP weights only. Backbone encoders remain frozen. The loop then returns to Phase 1, re-projecting all items with updated weights before the next HDBSCAN pass.

The training loop continues until the Adjusted Rand Index (ARI) is greater than 0.9 between the same samples in sequential iterations of the training loop. This indicates that the cluster assignments have become consistent and the model is no longer changing the organization of data in a meaningful way. After model convergence, the data will be organized into K clusters. The value of K is determined by the final pseudo-labeling step performed by HDBSCAN. The centroid for each of the K clusters is computed and passed to an LLM (Qwen 2.5-14B \cite{qwen2.5}) for a human-legible description.

For inference, a sample would be encoded by its modality backbone model, the embedding would then be passed through the MLP for its 256-d representation and compared to the cluster centroids using cosine similarity.

Cluster concept names are generated for human readability by taking the six items nearest to each cluster center using cosine similarity. Textual information for each item, the file name for image, audio and video samples and a 60-word content excerpt for text samples, is passed to a Qwen 2.5-14B model to extract a concept label. The human aesthetic register labels are withheld, allowing concept name assignment independent of the human category scheme.

\section{Results}

\begin{figure*}[t] 
  \centering
 \includegraphics[width=\textwidth]{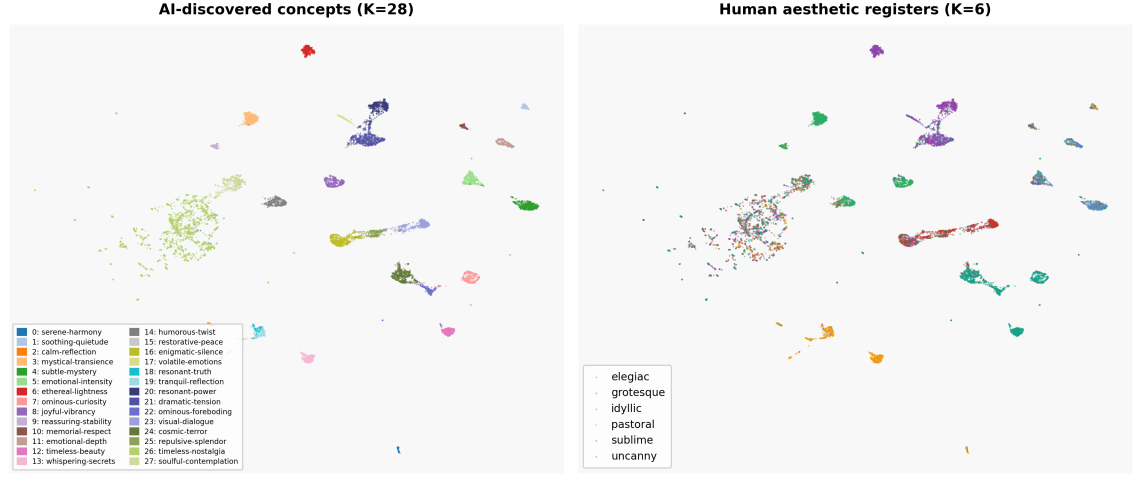} 
  \caption{UMAP Comparison of AI-discovered and Human Registry Embeddings Clusters, with 12,010 samples, using cosine metric. }
  \label{fig:umap}
  \Description{Two side by side cluster diagrams. The left side shows the 28 different clusters created by the AI-discovery categorization. The right side show the weakly assigned human label clusters. Both sides show realitve uniformity in cluster asignments except for an area of the left side graph showing a cluster with heterogeneous samples.}
\end{figure*}

The self-supervised multimodal clustering framework manages to organize the data into meaningful partitions without human supervision. Twenty-eight concept clusters are extracted from the data. The embedding UMAP \cite{mcinnes2018umap}, Figure~\ref{fig:umap}, shows a comparison of the AI clusters to the embedding clustering based on the six weakly assigned aesthetic labels. Both AI-discovered and human-aligned registers show distinct clusters. The AI-discovered clusters, given their finer granularity (28 vs six), are more homogeneous than the human-aligned register groups. 

Quantifying the divergence, we computed Normalized Mutual Information (NMI) to be 0.40, Adjusted Rand Index (ARI) is 0.15 and purity is 0.70 between AI-generated and human-aligned clusters. The NMI score shows that there is a moderate amount of shared information between AI-generated clustering and the human-aligned labels. The low ARI indicates that the AI is discovering a different structuring of the sample clusters while the high purity indicates the assignment is not arbitrary. 

\begin{figure}[h]
  \centering
 \includegraphics[width=\linewidth]{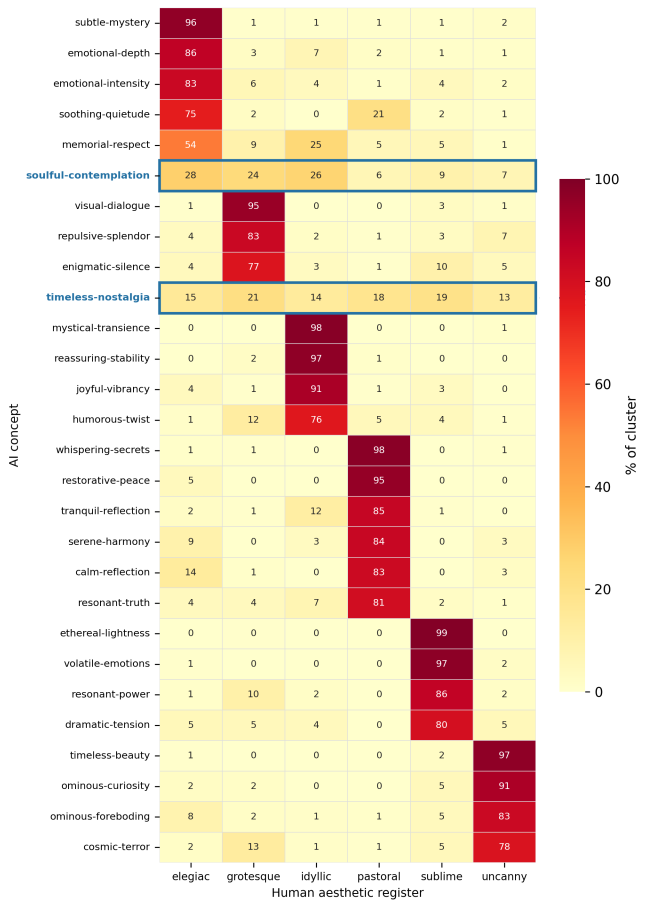}
  \caption{AI concept composition by human aesthetic register. Map is normalized by row and sorted by dominant register and cluster purity.}
  \label{fig:heatmap}
  \Description{Heat map with 28 rows and 6 columns comparing AI concept assignments and human-register labels. The map shows 26 of the 28 AI concepts align with 1 human register except 2 concepts: timeless-nostalgia and soulful-contemplation.}
\end{figure}

The far-left periphery contains a large cluster dominated by two AI-generated concept-labeled samples, Timeless-Nostalgia and Soulful-Contemplation, that stand out as cross-register exceptions. In the human-aligned graphic, this cluster is the most heterogeneous with sample representations from each of the assignable labels. 

\begin{figure}[h]
  \centering
 \includegraphics[width=\linewidth]{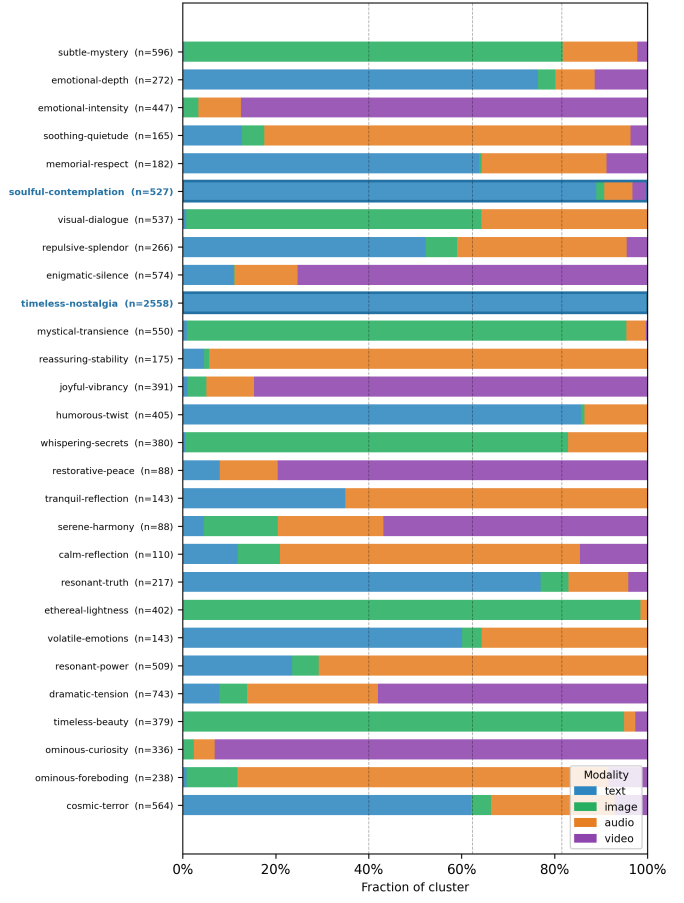}
  \caption{Modality composition of AI concepts clusters. Dashed lines indicate the dataset baseline.}
  \label{fig:modalities}
  \Description{Bar chart depicting the AI concept assignments based on modality. Timeless-nostalgia and Soulful-contemplation are both overwhelmingly represented by text.}
\end{figure}

The heat map presented in Figure~\ref{fig:heatmap} associates the AI-generated concepts with the human-aligned labels. The figure shows a tight overlapping of concepts between the AI and human labels, with the exception of Timeless-Nostalgia and Soulful-Contemplation. Figure~\ref{fig:modalities} shows the modal representation of samples assigned to each of the AI-generated clusters. Of the 28 converged clusters, 25 are anchored predominantly to a single modality, consistent with each medium carrying distinct perceptual properties, while three show more balanced modality representation. The cross-register exceptions are not cross-modal clusters but cross-human-register ones: Timeless-Nostalgia and Soulful-Contemplation draw from all six aesthetic registers while remaining text-dominated, revealing aesthetic dimensions that cut across emotional categories rather than across media boundaries. 

At iteration 1, before the training loop has refined the embedding space, HDBSCAN produces 22 modality-dominated clusters with a mean dominant-modality purity of 0.72. By convergence, the model reorganizes into 28 semantically coherent aesthetic clusters (mean purity 0.77), demonstrating that the cross-register structure captured by Timeless-Nostalgia and Soulful-Contemplation emerges through the iterative training rather than from backbone geometry. 

Soulful-Contemplation is primarily associated with elegiac, grote\-sque and idyllic. If we consider these on a valence-arousal scale we pass through three of the four quadrants with elegiac representing negative valence-low arousal, grotesque being negative valence-high arousal, and idyllic as positive valence-low arousal. This is interesting in comparison to the low representation of sublime, uncanny and pastoral. What is potentially being illustrated is the AI finding a different dimensionality of the data that was not captured in the human labels. The labels are intended to represent the emotions evoked by humans while consuming the media. Sublime is typically meant to be grand and overwhelming. Pastoral is focused on calm external and observational depictions. Uncanny seeks to foster detachment and other-worldliness. What was captured by the AI was a more internally motivated and reflective register, independent of the explicit emotion.

The AI-labeled, Timeless-Nostalgia samples represent a fairly even distribution of human labels. Queries for text samples focused on poems and prose, with sites like PoetryDB, Wikisource and Gutenberg being a large representation of the source. These sources are largely populated with public domain texts. This suggests that the utilization of older styles of language and writing is influencing the categorization when considering the contemporary data the large language models were trained on. What may be captured here is literary temporality. Since the public domain is composed of seminal and influential works, the AI labeled them as timeless. This bias in the data may also be responsible for the poor clustering on the human aesthetic assignments in Figure~\ref{fig:umap}.

\section{Conclusion}

We presented a self-supervised framework for multimodal aesthetic clustering that requires no cross-modal pairs or human supervision using a weakly labeled dataset spanning text, audio, image and video. The framework converges on 28 semantically coherent clusters, demonstrating that alignment across modalities can emerge from clustering signal alone. Divergence analysis (NMI=0.40, ARI=0.15, purity=0.70) reveals that AI-generated structure partially tracks human aesthetic registers while consistently discovering finer-grained and orthogonal organization. 

A key limitation is that human register labels were assigned at the collection level rather than per item, introducing noise into the divergence baseline. Additionally, collecting data from uncurated sources likely introduced bias, particularly in text, where public domain archives skew toward canonical Western literary works. Future work will prioritize curated artistic media collections, item-level human annotation to establish a cleaner ground truth, and human validation of AI-generated concept names. We also plan to evaluate the shared embedding space on downstream cross-modal retrieval tasks and extend the dataset across cultural traditions and at larger scale.  

\section*{Ethics and Privacy Statement}
This study strictly utilized publicly available data collected via an automated web tool designed for public data searches and sample retrieval. Because the data collection process was limited exclusively to publicly accessible archives and did not involve direct intervention, interaction, or solicitation with living human subjects, it did not require formal Institutional Review Board (IRB) oversight. 

To ensure compliance with data privacy principles and ethical research practices, the data collection methodology adhered strictly to the terms of service, robots.txt protocols, and user agreements of all sourced web platforms. No personally identifiable information (PII), proprietary private records, or restricted credentials were bypassed, scraped, or archived during the collection phase. All downloaded samples were processed, stored, and analyzed in an aggregated, de-identified format to fully mitigate any potential re-identification or privacy risks.

\bibliographystyle{ACM-Reference-Format}
\bibliography{references}

\end{document}